\documentstyle[aps,floats,epsf,twocolumn]{revtex}

\begin{document} \draft

\title{Mixed-state quasiparticle transport in high-$T_c$ cuprates}

\author{M. Franz}
\address{Department of Physics and Astronomy, Johns Hopkins University,
Baltimore, MD 21218
\\ {\rm(\today)}}
%
%\begin{abstract}
\address{~
\parbox{14cm}{\rm 
\medskip
Theory  of quasiparticle transport in the mixed
state of a $d$-wave superconductor is developed under the assumption
of disordered vortex array. A novel universal regime is identified
at fields above $H^*= c^*H_{c2}(T/T_c)^2$, characterized by a
{\em field-independent} longitudinal thermal conductivity $\kappa_{xx}^e$. 
It is argued that this behavior is responsible for the high-field plateau
in $\kappa_{xx}^e$ experimentally observed in high-$T_c$ cuprates.
}}
\medskip

%\end{abstract}
\maketitle

%\pacs{74.60.Ec,74.25.Fy,74.72.-h}

%
\narrowtext
Properties of the mixed state of high-$T_c$ cuprate superconductors keep 
surprising the scientific community in ways never imagined before. On one hand 
the collective behavior of vortices produces a multitude of ``vortex phases''
attributable to 
an intricate interplay between the thermal fluctuations, dimensional 
crossover, and pinning forces on vortices. At low 
temperatures, on the other hand, unexpected
and fascinating properties are observed 
related to the $d$-wave symmetry of the order
parameter and the consequent relativistic ``Dirac'' spectrum of 
the low-energy quasiparticle excitations. 

A recent 
example of such an unexpected behavior is the experimental observation
of the high-field plateau in the longitudinal thermal conductivity 
$\kappa_{xx}$ in 
Bi$_2$Sr$_2$CaCu$_2$O$_{8+\delta}$ (BiSCCO) and YBa$_2$Cu$_3$O$_{6+x}$ 
(YBCO) by Krishana, Ong and co-workers\cite{ong1,ong2} and by others
\cite{behnia1,zeini1,ando1}. 
In particular it is found that after an initial steep drop at low fields
(but well into the mixed state) $\kappa_{xx}$ becomes {\em field-independent}
above $H^*= c^*H_{c2}(T/T_c)^2$ and remains so up to the highest attainable 
fields $\sim 14$T. The implication is that both electronic and 
phononic contributions to $\kappa_{xx}$ separately become 
independent of field, with the initial drop attributed solely to electronic 
$\kappa_{xx}^e$\cite{ong2,zeini1}. These observations stand in a sharp 
contrast to the behavior of $\kappa_{xx}(H)$ found 
in conventional $s$-wave superconductors\cite{vinen1}, where vortices are 
strong scatterers of both electrons and phonons. Understanding
the physics of the plateau phenomenon therefore presents 
a considerable challenge to the theory of quasiparticle transport in cuprates. 
Its broader significance lies in the fact that simultaneous
measurement of the field-dependent longitudinal ($\kappa_{xx}$)
and transverse ($\kappa_{xy}$)
thermal conductivities, now feasible experimentally, contains a great wealth 
of information on the quasiparticle dynamics and in principle affords deduction
of inelastic scattering rate, thought to be important for the pairing 
mechanism in cuprates. 

The initial interpretation of the plateau involved a field-induced transition 
to a fully gapped state, such as $d_{x^2-y^2}+id_{xy}$ (or ``$d+id'$''),
which would effectively freeze out the quasiparticle
transport at low energies\cite{ong1}. Laughlin\cite{laughlin1} 
presented a compelling theoretical argument in support of such a scenario with
a weakly first order transition, but pointed out that opening of a gap with 
physically reasonable magnitude could not in itself account for the complete
suppression of the electronic transport
in BiSCCO at temperatures up to 30K. Furthermore, we find that sudden opening
of a sizeable gap leads to a {\em jump} in  $\kappa_{xx}(H)$, as opposed to a 
kink observed in \cite{ong1,behnia1}. Experimentally, there
appears to be little additional evidence for the 
$d+id'$ state, except perhaps for the apparent bound states
found by scanning tunneling microscopy (STM) in the vortex cores of YBCO
\cite{renner1} which should not exist in a pure $d_{x^2-y^2}$ state
\cite{franz1}. 

The existing theoretical treatments
\cite{laughlin1,simon1,hirschfeld1,ramakrishnan1}
so far side-stepped the important issue of the effect of magnetic field on
the quasiparticle mean free path (MFP), determined primarily by scattering of 
quasiparticles by the Abrikosov vortices. Detailed theory of the MFP due to 
vortices in
$s$-wave superconductors was developed long ago by Cleary\cite{cleary1}.
The $d$-wave problem, however, is fundamentally different\cite{franz1}
and, except for heuristic treatments given
in Refs.\ \cite{ong3,solomon1}, no analogous theory exists at present.
Here we develop such a theory and 
show that field-independent longitudinal  
thermal conductivity arises quite naturally in a pure $d_{x^2-y^2}$ state
above a crossover field $H^*(T)$. This behavior reflects
an exact compensation between
the enhancement of the quasiparticle density of states (DOS) in the presence 
of non-uniform
superflow (the ``Volovik effect'') \cite{volovik1} and the concomitant 
reduction in the quasiparticle mean free path $\ell$ due to increased
scattering from vortices  
in a disordered vortex array. The limiting high-field value of 
$\kappa_{xx}^e(H)$ is universal in the similar sense as the impurity induced
universal microwave conductivity $\sigma(\omega\to 0)$ predicted by Lee
\cite{lee1}. The approach to this limiting value depends on the distribution 
of vortices in the sample and will therefore be material and sample 
dependent.

The basic physics of the new universal regime can be understood from the 
following simple argument. 
In the elementary Boltzmann-type treatment the electronic thermal conductivity 
of a normal metal is given by 
\begin{equation}
\kappa_{xx}^e={1\over 3}v_Fc_v\ell,
\label{bk}
\end{equation}
where $c_v$ is the electronic specific heat and $\ell$ is the MFP. 
We now argue that this relation remains valid in the mixed state of 
a $d$-wave superconductor at fields above $H^*$.
As first pointed out by Volovik\cite{volovik1}, the
superfluid velocity field around vortices induces a Doppler shift in the 
excitation spectrum of quasiparticles,
which in turn leads to a finite DOS at the Fermi level. For $H\gtrsim H^*$ the
system behaves effectively like a normal metal and we expect (\ref{bk})
to hold. The residual DOS scales as $N_H(0)\sim\sqrt{H}$ 
and gives rise to the well known low temperature specific heat\cite{volovik1}
\begin{equation}
c_v\simeq k_0N_FT\sqrt{H/H_{c2}}
\label{cv}
\end{equation}
where $N_F$ is the normal-state DOS, and
$k_0$ is a constant of order unity. This type of behavior is indeed
observed experimentally\cite{moler1}.
 
We now estimate the vortex contribution $\ell_H$ to the 
total MFP $\ell^{-1}=\ell_0^{-1}+\ell_H^{-1}$.
The central assumption we make is that in the 
regime of experimental interest the vortex
lattice is {\em disordered} in the sense that it possesses no 
long range translational
order, and will therefore scatter quasiparticles\cite{remark1}.
At low energies $\ell_H$ is dominated by 
the quasiparticle scattering from the superfluid velocity field 
${\bf v}_s({\bf r})$. Scattering from  the vortex cores is down by a factor
$(\xi/a_v)^2=H/H_{c2}\ll 1$\cite{franz2}, where $a_v=\xi\sqrt{H_{c2}/H}$ is the
average inter-vortex distance,  $\xi=v_F/\pi\Delta_0$ 
is the coherence length, and $\Delta_0$ is the maximum gap.
  In the (disordered) 
vortex lattice ${\bf v}_s({\bf r})$ varies on the length scale set by $a_v$. 
Since $a_v$ is the only relevant length scale in the problem\cite{remark7}, 
one expects on general grounds that
\begin{equation}
\ell_H\simeq k_1 a_v\propto \sqrt{H_{c2}/H},
\label{ell1}
\end{equation}
where $k_1$ is a field-independent constant. 
This conclusion is indeed confirmed by an explicit calculation of the
quasiparticle propagator outlined below, as well as by a
calculation of the vortex transport scattering cross-section $\sigma_{\rm tr}$
\cite{franz2}, carried out along the lines of the classical treatment by Cleary
\cite{cleary1}. We note that Eq.\ (\ref{ell1}) is also consistent with the 
result of Ref.\cite{solomon1} based on a heuristic argument involving the
Andreev reflection.   

In sufficiently strong fields Eq.\ (\ref{ell1}) implies that
$\ell_H\ll\ell_0$\cite{remark2} and 
the total MFP will be dominated by vortices: $\ell\approx\ell_H$. 
On substituting (\ref{cv}) and (\ref{ell1}) into (\ref{bk}) we 
arrive at the desired result that for 
$H\gg H^*$, $\kappa_{xx}^e$ will approach the {\em field-independent}
 universal value
\begin{equation}
\kappa_{xx}^{eH}/T=k'\pi N_Fv_F^2/3\Delta_0,
\label{ku}
\end{equation}
with $k'=k_0k_1$. Evaluation 
of $k'$ below shows that $\kappa_{xx}^{eH}$ is identical 
to the ``universal'' thermal conductivity $\kappa_{00}^e$ due to impurities
predicted by various authors\cite{hirschfeld1,graf1} and observed 
experimentally by Taillefer and co-workers\cite{taillefer1}. 

The argument based on Eq.\ (\ref{bk}) captures the essential physics of the 
new universal regime and provides a very natural explanation of the plateau
phenomenon observed in cuprates\cite{ong1,ong2,behnia1}
in terms of fundamental properties of the Dirac fermions.  We now present
a more rigorous treatment of $\kappa_{xx}^e(H)$, 
based on the Kubo formula for the
heat current response. Besides supplying the unknown constant $k'$, such 
calculation provides the necessary confidence in our result (\ref{ku}) and 
helps understanding the approach to the universal limit with the increasing 
$H$.

In the absence of field the quasiparticle propagator is a $2\times 2$ matrix
in the Nambu space (taking $\hbar=1$)
\begin{equation}
\hat G_0(\omega,{\bf k})={\omega+\hat\tau_3\epsilon_{\bf k}+
\hat\tau_1\Delta_{\bf k} \over
\omega^2-\epsilon_{\bf k}^2-\Delta_{\bf k}^2},
\label{g0}
\end{equation}
where $\hat\tau_i$ are the Pauli matrices. In intermediate magnetic fields the
coupling of quasiparticles to the superflow around the  
vortices is well described by the semiclassical replacement
$\omega\to\omega-{\bf k}\cdot{\bf v}_s({\bf r})$ in Eq.\ (\ref{g0}) 
\cite{hirschfeld1,franz3}. 
In a simple London model, which will be sufficient for our purposes, 
the superfluid velocity field is given by\cite{london1}
\begin{equation}
{\bf v}_s({\bf r})={\pi\lambda^2\over m}\int {d^2k\over(2\pi)^2}
{i{\bf k}\times\hat z\over 1+\lambda^2k^2}\sum_ie^{i{\bf k}\cdot({\bf r}-
{\bf R}_i)}.
\label{vs}
\end{equation}
Here $\lambda$ is the London penetration depth, $\hat z$ is a unit
vector along the field direction, and $\{{\bf R}_i\}$ denotes vortex 
positions. 
In a disordered vortex array, on the length scales large compared
to $a_v$, propagation of quasiparticles will be described by the Greens
function {\em averaged} over $\{{\bf R}_i\}$:
\begin{equation}
\hat G(\omega,{\bf k})=\langle \hat G_0(\omega-{\bf k}\cdot{\bf v}_s,{\bf k}) 
\rangle_{\{{\bf R}_i\}}.
\label{g1}
\end{equation}
If we now define a probability density
\begin{equation}
{\cal P}(\eta)=\langle \delta[\eta-{\bf k}\cdot{\bf v}_s({\bf r})] 
\rangle_{\{{\bf R}_i\}},
\label{p1}
\end{equation}
we may rewrite (\ref{g1}) as 
\begin{equation}
\hat G(\omega,{\bf k})=\int d\eta {\cal P}(\eta)\hat G_0(\omega-\eta,{\bf k}),
\label{g2}
\end{equation}
and all the information on the vortex array is now encoded in ${\cal P}(\eta)$.
%%%%%In order to evaluate this quantity an assumption about the distribution
%of $\{{\bf R}_i\}$ must be made. 
Making the most natural ansatz that
the vortex positions are random and uncorrelated, it can be shown
\cite{solomon1} that 
${\cal P}(\eta)=(2\pi\sigma_H^2)^{-1/2}e^{-\eta^2/2\sigma_H^2}$  with
\begin{equation}
\sigma_H^2=k_\alpha k_\beta\langle v_s^\alpha({\bf r}) v_s^\beta({\bf r})
\rangle_{\{{\bf R}_i\}}\simeq
{\pi^2\over8}\left(H\over H_{c2}\right)\Delta_0^2\ln\kappa.
\label{sig}
\end{equation}
Here $\kappa=\lambda/\xi$, 
and the last equality follows from Eq.\ (\ref{vs}).

Evaluation of the propagator $\hat G$ is somewhat 
complicated by the fact that the 
$\eta$-integral in Eq.\ (\ref{g2}) cannot be expressed in terms of
elementary functions. This difficulty can be avoided by taking ${\cal P}$ to 
be a Lorentzian ${\cal P}(\eta)=\pi^{-1}\sigma_H/(\eta^2+\sigma_H^2)$ 
rather than
a Gaussian, in which case the integration is elementary and one obtains
\begin{equation}
\hat G(\omega,{\bf k})=\hat G_0(\omega-i\sigma_H,{\bf k}).
\label{g3}
\end{equation}
Scattering from vortices therefore 
results in a self-energy correction to the Greens
function,  which, for the particular case
of Lorentzian distribution, is simply a constant 
$\Sigma_H(\omega,{\bf k})=i\sigma_H$. Had we kept the Gaussian 
distribution, the self energy would be similar at low energies, but would tend
to zero for $|\omega^2-\epsilon_{\bf k}^2-\Delta_{\bf k}^2|\gg\sigma_H^2$. 
This additional structure results in slight mathematical complications 
but does not change the result for $\kappa_{xx}^e(H)$ qualitatively. We 
therefore choose to proceed with the simpler form (\ref{g3}) and treat it 
either as a reasonable low energy approximation to Eq.\ (\ref{g2}) or as
a slightly different physical problem with a vortex distribution 
$\{{\bf R}'_i\}$ resulting in the Lorentzian distribution ${\cal P}(\eta)$. 

Eq.\ (\ref{g3}) may be used to calculate the field-induced
DOS, $N_H(\omega)=-(2\pi)^{-1}\sum_{\bf k} {\rm Tr}\hat A(\omega,{\bf k})$, 
where  $\hat A(\omega,{\bf k})=2{\rm Im}\hat G(\omega,{\bf k})$ is the
spectral function.
One obtains $N_H(0)\propto\sigma_H\propto\sqrt{H}$, in agreement with
Volovik's result\cite{volovik1}. Interpretation of $\sigma_H$ as
the scattering rate due to vortices allows estimation of the vortex MFP 
$\ell_H=\bar v/\sigma_H$, where $\bar v$ is the average quasiparticle velocity.
Taking $\bar v= \sqrt{v_Fv_\Delta}$ \cite{simon1} we find 
$\ell_H=a_v(8v_\Delta/v_F\ln\kappa)^{1/2}\simeq 0.5a_v$ (for 
$v_F/v_\Delta=7$ and $\kappa=70$), in agreement with our naive estimate 
(\ref{ell1}). We finally note that since the scattering rate $\sigma_H$ 
is proportional to $\sim\sqrt{n_v}$ (with $n_v=H/\Phi_0$ the vortex density), 
Eq.\ (\ref{g3})  will be in 
general difficult to derive from conventional diagrammatic techniques
which typically employ low-density approximation and therefore yield
scattering rates proportional to $n_v$. Physically, the $\sqrt{n_v}$ 
non-analyticity reflects the long ranged $1/r$ nature of the ${\bf v}_s$
field associated with a single vortex.

We are now in the position to evaluate the thermal conductivity tensor from 
the standard linear response theory. Following Ambegaokar and Tewordt
\cite{amb1}, we find for a $d$-wave superconductor in 2D,
\begin{equation}
{\kappa_{ij}^e\over T}={1\over 32\pi m^2}\int_{-\infty}
^\infty {d\omega\over T}\left({\omega\over T}\right)^2{\rm sech}^2
\left({\omega\over 2T}\right) K_{ij}(\omega),
\label{k1}
\end{equation}
where (neglecting vertex corrections)
\begin{equation}
K_{ij}(\omega)=\int {d^2k\over (2\pi)^{2}} k_ik_j 
{\rm Tr}\{\hat\tau_3\hat A(\omega,{\bf k})\hat\tau_3\hat A(\omega,{\bf k})\}.
\label{ker1}
\end{equation}
At temperatures low compared to $\Delta_0$ it is permissible to linearize
the quasiparticle spectrum in the Dirac cones near the four nodes of the gap 
function $\Delta_{\bf k}=\Delta_0\cos(2\theta)$\cite{simon1,lee1}. 
Taking $\epsilon_{\bf k}=v_F k_1$ and $\Delta_{\bf k}=v_\Delta k_2$ in the 
new coordinate system $(k_1,k_2)$ with the origin at the node, one can 
explicitly perform the integral in (\ref{ker1}) to obtain
\begin{equation}
K_{xx}(\omega)={4k_F^2\over \pi v_Fv_\Delta}
\left[1+\left({\omega\over\sigma}+{\sigma\over\omega}\right)
\arctan{\omega\over\sigma}\right].
\label{ker2}
\end{equation}
Here $v_\Delta=2\Delta_0/k_F$ and we have replaced $\sigma_H$ by 
 the total scattering
rate $\sigma=\sigma_0+\sigma_H$, with $\sigma_0$ describing both elastic and 
inelastic processes in the superconductor at $H=0$. The proper
description of these processes would presumably require a self energy 
$\Sigma_0(\omega,{\bf k})$ which depends strongly on both of its arguments. 
However, absent the detailed microscopic theory of such processes,
we follow\cite{hirschfeld3} and simply model their effect by a 
phenomenological constant $\sigma_0$, 
which we interpret as a process-specific average of $\Sigma_0(\omega,{\bf k})$.

For arbitrary $\sigma$ and $T$ the  expression (\ref{k1}) 
for $\kappa_{xx}^e$ must be evaluated numerically. However, the leading 
terms may be readily obtained in a perturbative expansion.
%by noting that the integral is dominated by values of $K_{xx}(\omega)$ near
%$\omega=2T$. Expanding the $\arctan$ in (\ref{ker2}) at small and large 
%arguments 
We obtain, for $\sigma\gg 2T$ (high field regime):
\begin{equation}
{\kappa_{xx}^e\over T}\simeq {\pi v_F^2 N_F\over 6\Delta_0}
\left( 1+{7\pi^2\over 15}{T^2\over \sigma^2}\right);
\label{hi}
\end{equation}
and for $\sigma\ll 2T$ (low field regime):
\begin{equation}
{\kappa_{xx}^e\over T}\simeq {v_F^2 N_F\over 2\Delta_0}
\left[{9\zeta(3)\over 4}{T\over\sigma}+{\ln2\over2}{\sigma\over T}\right]. 
\label{lo}
\end{equation}
The first term in Eq.\ (\ref{hi}) reproduces Eq.\ (\ref{ku}) for $k'=1/2$ and
represents 
the universal thermal conductivity $\kappa_{xx}^{eH}\equiv\kappa_{00}^e$
\cite{hirschfeld1,graf1}, which does not depend on the details
of the vortex distribution, as long as there is no long range order. 
This result justifies the usage of the simple Boltzmann approach (\ref{bk})
in the mixed state of a $d$-wave superconductor. The
leading correction in (\ref{hi}) is non-universal; e.g. one may show that 
the power will change to $(T/\sigma)^4$ for a Gaussian ${\cal P}$. 

Eqs.\ (\ref{hi}) and (\ref{lo}) provide a simple tool for extracting the
zero-field scattering rate $\sigma_0$ from the experimental data. Assuming 
that the phonon contribution $\kappa_{xx}^p$ to the thermal 
conductivity $\kappa_{xx}$ is field-independent\cite{ong2,zeini1},
the total drop in $\kappa_{xx}(H)$ between
$H=0$ and the plateau is simply related to $\sigma_0$ by
\begin{equation}
\sigma_0=
\biggl\{ \begin{array}{ll}
c_1T(1+\delta\kappa_{xx})^{-1}, \ \ \ & \sigma_0\ll T \\
c_2T(\delta\kappa_{xx})^{-1/2}, &       \sigma_0\gg T
\end{array}
\label{drop}
\end{equation}
Here $\delta\kappa_{xx}=[\kappa_{xx}(0)-\kappa_{xx}(\infty)]/\kappa^e_{00}$,
$\kappa_{xx}(\infty)$ denotes the plateau value, 
$c_1=27\zeta(3)/4\pi\simeq 2.58$ and $c_2=\sqrt{7\pi^2/15}\simeq 2.15$.
Applying these relations to the data on underdoped YBCO
\cite{ong2}  and using $\kappa_{00}^e/T=0.019$W/K$^2$m \cite{taillefer1}, we
obtain $\sigma_0\approx 0.4$meV at 10K, increasing to about 7meV at 50K
and 19meV at 60K (just below $T_c=63$K).

\begin{figure}[t]
\epsfxsize=8.5cm
\epsffile{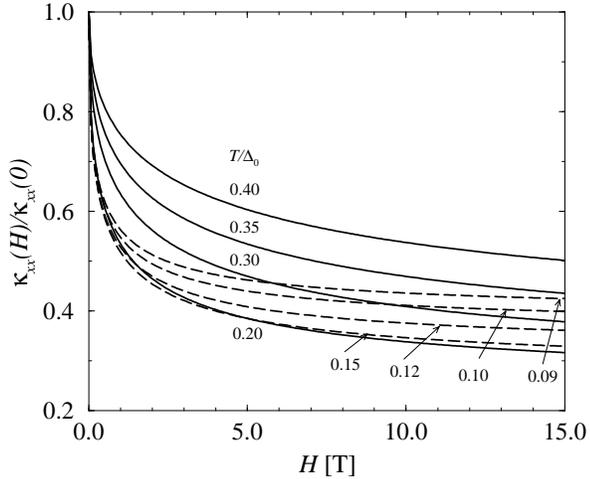}
\caption[]{Normalized electronic thermal conductivity $\kappa_{xx}^e(H)$
as a function of $H$ at selected temperatures as 
calculated from Eq.\ (\ref{k1}). Parameters used are $H_{c2}=160$T, 
$\kappa=70$, $\gamma_0=0.1$ and $\gamma_3=4$. }
\label{fig1}
\end{figure}

To illustrate the behavior of $\kappa_{xx}^e(H)$ over the wide range of
fields and temperatures in Fig.\ \ref{fig1} 
we have evaluated Eq.\ (\ref{k1}) with kernel 
(\ref{ker2}) numerically, adopting a phenomenological expression for the 
scattering rate $\sigma_0/\Delta_0=\gamma_0+\gamma_3(T/\Delta_0)^3$. Here
$\gamma_0$ represents the residual scattering rate due to impurities and 
the $\gamma_3$ term models the inelastic scattering rate\cite{hirschfeld3}
 which is known to collapse rapidly below $T_c$\cite{ong3}. 
Similarity of Fig.\ \ref{fig1} to the experimental data on the underdoped
YBCO \cite{ong2}
is striking: the positive curvature of $\kappa_{xx}(H)$ is enhanced with 
decreasing $T$ and the curves approach a field-independent value at high
fields. The characteristic crossing of the curves near 
$T/\Delta_0 \simeq 0.15$, which reflects the crossover in $\sigma_0(T)$
from inelastic scattering at high $T$ (solid lines) to elastic scattering 
at low $T$ (dashed lines)  is also present. 

In some samples of BiSCCO, the plateau is reached abruptly, with a 
discontinuity in the 
slope of $\kappa_{xx}(H)$ at $H^*$\cite{ong1,behnia1,ando1}, indicative of a 
phase transition. This ``kink'' behavior is not reproduced by the present 
simple model, which, however, can be presumably appropriately modified once
the nature of the phase transition is understood. We take a position here
that it is the plateau phenomenon which is universal and significant (since
it is observed in more than one compound) while the approach to it is a
material-specific issue of secondary importance. 

Aubin and co-workers \cite{behnia1} further report that the behavior of 
$\kappa_{xx}(H)$ in BiSCCO qualitatively changes below 1K, in that it
{\em increases} with magnetic field, approximately as $\kappa_{xx}(H)\sim
\sqrt{H}$. From Eq.\ (\ref{bk}) the most natural interpretation is that,
below 1K, $\ell$ becomes independent of $H$. There are two possible reasons
for this. The vortex array may order at low $T$, in which case the
Bloch theorem prevents quasiparticles from being scattered by vortices
\cite{hirschfeld1}. The second possibility stems from
the computation of the transport scattering cross-section of a $d$-wave 
vortex\cite{franz2}, which indicates that $\sigma_{\rm tr}(\omega)$ vanishes
as $\omega^{3}$ for quasiparticle energies 
$\omega\lesssim\Delta_0/(k_F\lambda)$. Thus, at low energies, vortices
become ``transparent'', and even disordered vortex array 
will not scatter quasiparticles. 

An immediate consequence of our picture is that {\em all} of the high-$T_c$ 
compounds with the $d_{x^2-y^2}$ gap should exhibit the plateau phenomenon
with the universal value $\kappa_{xx}^{eH}/T=\pi v_F^2 N_F/6\Delta_0$
in sufficiently high fields. We predict that the existence of
plateau will correlate with the absence of long range order in the vortex 
array, and will fade away at the lowest temperatures. 
The formalism 
developed here also permits calculation of the Hall conductivity 
$\kappa_{xy}^e$. However, this constitutes a more complicated problem since
one has to consider the particle-hole asymmetry in the vortex scattering 
rate\cite{cleary1} as well as corrections to the linearized Dirac spectrum
\cite{simon1}.

The author is indebted to N. P. Ong for sharing his insights 
and his data prior to publication; and to Y. Ando, A. V. Balatsky, K. Behnia,
A. J. Berlinsky, 
P. J. Hirschfeld, C. Kallin, A. J. Millis, L. Taillefer and Z. Te\v{s}anovi\'c
for valuable discussions. This work was supported by NSF grant DMR-9415549 
and was done in part at the Aspen Center for Physics.


\begin{references}
\bibitem{remark1}  Absence of the neutron 
diffraction peaks above 60mT in BiSCCO\cite{cubitt1} and the real space vortex
images obtained by STM\cite{renner1,renner2} clearly support this assumption. 
The finding that $\kappa_{xx}$ initially decreases with increasing $H$
\cite{ong1,ong2,behnia1,zeini1,ando1} is also most naturally explained by this 
assumption.
\bibitem{remark2} This condition is easily satisfied in clean single 
crystals where $\ell_0$ typically exceeds several thousands \AA \cite{ong3}.
\end{references}
\end{document}